\documentstyle[11pt,newpasp,twoside,epsf]{article}
\def\edcomment#1{\iffalse\marginpar{\raggedright\sl#1\/}\else\relax\fi}
\reversemarginpar

\begin{document}
\title{The relation between galactic properties and cluster structure}
\author{Bianca M. Poggianti}
\affil{Osservatorio Astronomico di Padova-INAF, Italy}
\author{Terry J.\ Bridges}
\affil{Anglo-Australian Observatory, Australia}
\author{M. Yagi}
\affil{National Astronomical Observatory, Mitaka, Tokyo, Japan}
\author{Y. Komiyama}
\affil{Department of Astronomy, University of Tokyo, Japan}
\author{Dave Carter}
\affil{Liverpool John Moores University, Birkenhead, Wirral, UK}
\author{Bahram Mobasher}
\affil{Space Telescope Science Institute, Baltimore, USA}
\author{S. Okamura}
\affil{Department of Astronomy, University of Tokyo, Japan}
\author{N. Kashikawa}
\affil{National Astronomical Observatory, Mitaka, Tokyo, Japan}

\begin{abstract}
A satisfactory understanding of the origin of the dependence of galaxy
properties on their environment has remained, so far, out of reach.
In the light of numerous observational results and substantial
theoretical progress obtained for clusters of galaxies in the last
years, a primary goal is to understand how the star formation activity
depends on cluster substructure, i.e. on the merging/accretion history
of a cluster.  In this contribution we present a case in which it is
possible to identify the cluster environment, and in particular the
intracluster medium and the recent infall history of galaxies onto the
cluster, as the cause for an abrupt change in the star formation
histories of a subset of galaxies in the Coma cluster.
\end{abstract}

\section{Introduction}

Fundamental galaxy properties such as star formation history,
morphology and gas content vary in a systematic way from dense
environments like clusters of galaxies to low density regions in
the general field.  Similarly, it has long been known that
there are systematic variations of {\it galaxy}
properties with {\it cluster} properties.

Already in 1974, on a sample of 15 clusters, Oemler tentatively
divided clusters into three types. He noted that clusters with cD
galaxies and rich in ellipticals have a smooth spherical distribution
and are centrally concentrated, while clusters that are dominated by
S0s are not as smooth, dense or centrally concentrated as cD clusters,
and clusters of irregular appearance are dominated by spiral,
star-forming galaxies.  A few years later, in 1977, Bahcall first
showed that the fraction of spiral galaxies in a cluster decreases
rapidly with increasing X-ray luminosity (see also Edge \& Stewart
1991).

Subsequent studies on the correlation between galaxy properties and
global properties of clusters have been innumerable, especially
searching for correlations between the star formation activity in
galaxies and the cluster X-ray luminosity (Smail et al. 1998,
Ellingson et al. 2001, Balogh et al. 2002a), the cluster richness
(Margoniner et al. 2001) and the presence of substructure (Wang \&
Ulmer 1997, Metevier et al. 2000, Pimbblet et al. 2002).  At some
level the global cluster properties (X-ray and optical luminosities,
richness, concentration etc.) are all correlated with each other,
though with a large scatter, and it is hard to pin down what exactly
is the primary correlation with galaxy properties that drives all the
others.

A central question is surely how the star formation activity depends
on cluster substructure, i.e. on the merging/accretion history of the
cluster.  There are some very good examples of cluster-cluster mergers
where the star formation activity in galaxies, as measured by radio
continuum emission, is exceptionally high, such as Abell 2125
(Dwarakanath \& Owen 1999) and Abell 2255 (Miller \& Owen 2003). In
principle it is not justified to inequivocally attribute this activity
to an ``enhancement'' of star formation due to the cluster-cluster
merging: the activity observed could simply reflect an higher average
activity in the substructures that are present in the cluster, given
that the average star formation is known to be higher in lower
density/less massive environment. Thus, the cause of the observed
activity needs to be investigated on a one-by-one case. In Abell 2255,
for example, the alignment of the star-forming galaxies along a
preferential direction, perpendicular to the merger axis, strongly
supports the hypothesis that the bursts of star formation were indeed
triggered by the cluster-cluster merger, induced by ram pressure
variations at the shock front (Miller \& Owen 2003).  On the other
hand, there are also merging clusters where only a low star formation
activity is detected, suggesting that the gas reservoir in the galaxy
populations of the two merging clusters must also play a role (Miller
2003). The dynamical state of the cluster and the type of galaxies
that happen to be falling in at the time we observe it are the two
crucial factors also for understanding the gas (HI) content of
galaxies in clusters (van Gorkom 2003).

Galaxy properties are
also known to be closely linked with their {\it local} environment,
regardless of the ``global'' environment.  The most striking evidence
for this is the morphology-density relation (Dressler 1980, Dressler
et al. 1997), paralleled by a star-formation/density relation (Lewis
et al. 2002, Gomez et al. 2003).  The higher incidence of ellipticals
in regions of high local density, and the corresponding preference for
spirals to reside in low density regions is known to be valid across a
wide range of ``global'' environments, for both concentrated and
irregular clusters, for rich and poor systems (Dressler 1980, Postman
\& Geller 1984), though an additional second-order effect related to
the {\it global} environment seems to be present, in the sense that at
a given local density more massive and relaxed clusters show a small
excess of early-type galaxies compared to less massive and irregular
clusters (Dressler 1980, Balogh et al. 2002b).  One of the open
questions is therefore to what extent are the differences in galaxy
properties in various types of clusters due to the fact that the {\it
average local} environment varies as a function of the cluster type.

The findings cited above, together with many others in the literature,
prove that both {\it local} 
and {\it global} effects are at work. In this talk we present the case of 
a cluster-related phenomenon, where it is possible to identify
the cluster environment, and in particular the intracluster medium
and the recent infall history of galaxies onto the cluster, 
as the cause for an abrupt change in the star formation histories
of a subset of the Coma galaxies.

\section{Substructure and post-starburst galaxies in the Coma cluster}

In a recent spectroscopic survey of galaxies in the Coma cluster
(Mobasher et al. 2001), we found that a significant fraction of the
cluster dwarf galaxy population 
has post-starburst/post-starforming spectra (Poggianti et al. 2003,
hereafter P03).  This type of spectrum (``k+a'', or ``E+A'') has no
emission-line detected and equivalent width
EW($\rm H\delta)> 3$ \AA, indicating a
galaxy with no current star formation activity which was forming stars
at a vigorous rate in the recent past (last 1.5 Gyr).

Numerous spectroscopic surveys of galaxies in distant clusters have
found significant populations of luminous k+a galaxies (see for
example Couch \& Sharples 1987, Abraham et al. 1996, Dressler \& Gunn
1992, Fisher et al. 1998, Dressler et al. 1999, Tran et al. 2003, but
see also Balogh et al. 1999).  However, k+a galaxies as luminous as
those in distant clusters ($M_V \leq -20$) are absent in Coma,
where k+a's are detected at magnitudes typically fainter than $M_V \sim -18$
(P03).
The different luminosity distributions of k+a galaxies in clusters at
$z\sim 0.5$ and in Coma most likely reflects an evolution in the properties
of the infalling galaxies, and provides further evidence of a
``downsizing effect'' (P03, see also Tran et al. 2003): the maximum
luminosity/mass of galaxies with significant star formation activity
seems to decrease at lower redshifts.
In Coma we are observing late-type starforming galaxies becoming dwarf 
spheroidals (P03), while the descendants of k+a's at high redshift
will be among the most massive early-type galaxies today (Tran et al. 2003).

In the color-magnitude diagram, a group of blue and a group of red
k+a's can be easily distinguished in Coma.  The average EW($\rm
H\delta$) of the blue group is significantly stronger than that of the
red group.  The blue, strong k+a's most likely correspond to ``young''
k+a's (observed soon after the termination of star formation, $<$ 
300 Myr) and the
red, weaker k+a's are ``old'' ones (observed at a later stage of the
evolution, 0.5-1.5 Gyr).  
Most of the blue k+a's, and a few of the red k+a's, have
EWs stronger than 5 \AA $\,$, testifying that a starburst occurred in
the galaxy before star formation was quenched.
Note that Balmer-enhanced galaxies in Coma, their luminosities and positions
within the cluster have been the subject of several works
by Caldwell, Rose and collaborators (e.g. Caldwell et al. 1993, see
P03 for all references). 

\begin{figure}[h]
\plotone{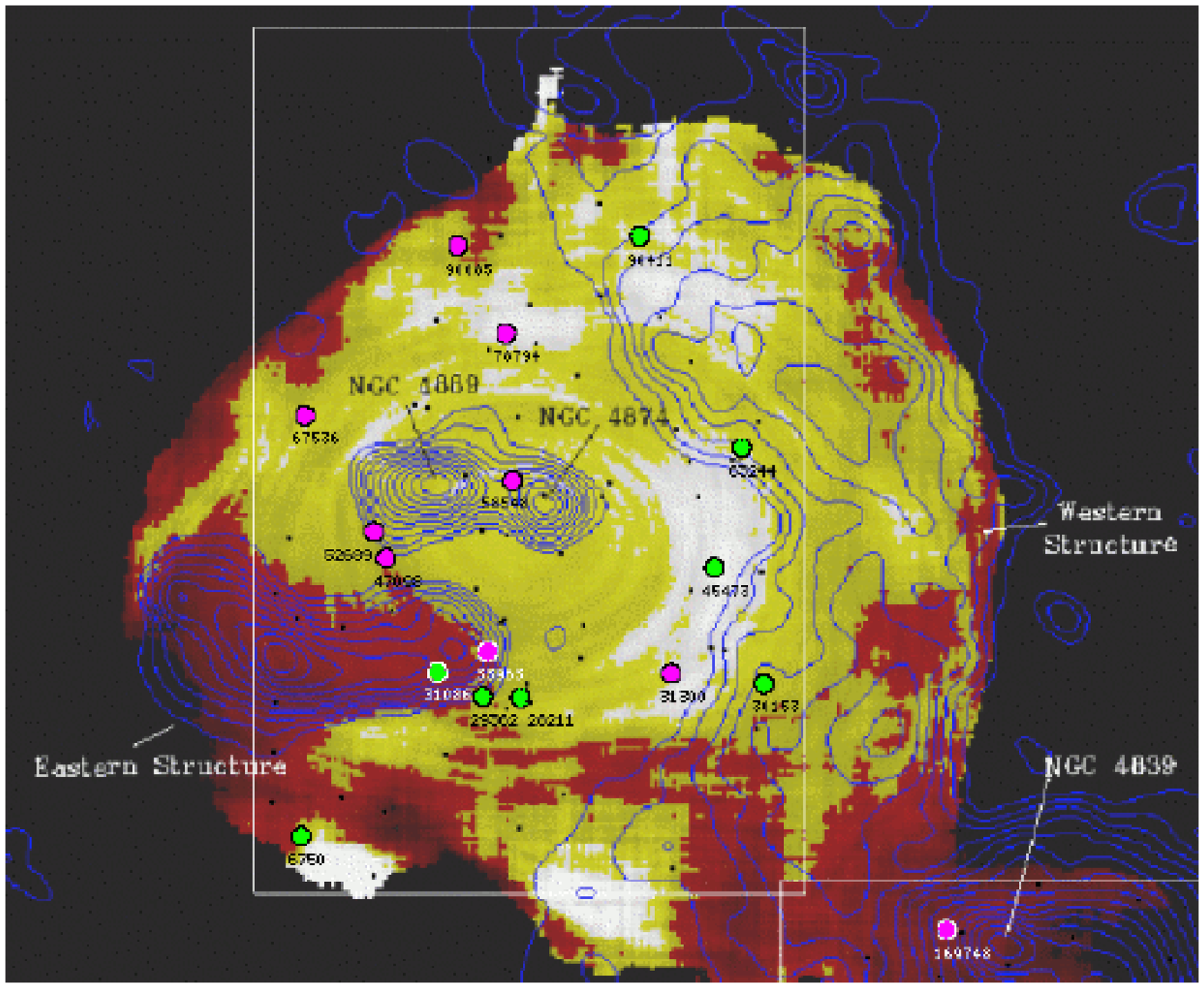}
\caption{{\it N.B. This plot is best viewed in color.}
Position of k+a galaxies with respect to X-ray substructure
and X-ray temperature map. Only the central field of Coma is shown
here, see P03 for a full map. Strong-lined k+a's with EW($\rm
H\delta)>5$ \AA $\,$ are shown as green dots, while weaker k+a's are
plotted as magenta dots.  Tiny black dots are dwarf Coma members with
velocities $> 7200 \, \rm km \, s^{-1}$.  X-ray residuals from
Neumann et al. (2003) are plotted as contours and clearly identify two
substructures (Western and Eastern substructures), in addition to the
NGC4839 peak in the South-West and the excess of emission towards the
two central galaxies (NGC4874 and NGC4889).  The lowest contour and
the step width between two contours are each 5 $\sigma$.  The hardness
ratio image (2-5 keV/0.5-2keV, Neumann et al. 2003) is shown in color. Red
regions correspond to temperatures below 8 keV, yellow to $kT>8$ keV
and white regions to $kT>10$ keV.  The rectangles show the limits of
the two fields of our photometric and spectroscopic survey (Coma1
towards the cluster center and Coma3 in the South-West). Each
rectangle is about 1 by 1.5 Mpc.}
\end{figure}

A suggestive clue about the possible physical mechanism responsible
for the k+a spectra comes from the recent X-ray mosaic observations of
Coma obtained with {\it XMM-Newton}. Coma has two central dominant
galaxies, NGC 4874 (a cD galaxy) and NGC4889 (a very bright elliptical), 
and another cD galaxy, NGC4839,
that dominates a substructure South-West of the center (Fig.~1).  Neumann
et al. (2003) have recently identified and discussed X-ray
substructure by fitting a smooth profile and subtracting it from the
data. The residuals reveal several structures, that are shown as
contours in Fig.~1: besides the well known NGC4839 South-West group,
Neumann et al. identify a large residual to the West of the cluster
centre (``Western structure'' in Fig.~1) elongated along the
North-South direction, and a filament-like structure South-East of the
centre (``Eastern structure'' in Fig.~1), elongated along the
East-West direction.  The temperature map shown in color in Fig.~1
sheds further light on the accretion history of Coma. Neumann et
al. conclude that the region of high temperature observed between the
Western structure and the Coma center is caused by the infall of this
structure, either via compression or via shock waves.  These authors
consider the two maxima in the western structure to be likely the
result of the disruption of a galaxy group during its infall, instead
of two galaxy groups falling at the same time.  In contrast, the
South-Eastern structure is cooler than the mean cluster temperature
and is associated with a low-mass galaxy group dominated by two large
galaxies, NGC4911 and NGC4921.  Based on the filamentary form of this
structure, the same authors conclude it is observed during the infall
process while being affected by ram pressure stripping close to the
cluster centre.

The coincidence of the position of the strongest k+a galaxies and the
X-ray structures is striking. Four k+a's with EW($\rm H\delta)>5$ \AA
$\,$ (green dots in Fig.~1) trace the edge of the Western structure
towards the Coma centre. Another three are associated with the
Eastern structure, all at its western boundary.  
Thus, young post-starbursts are distributed
close to the edge of infalling substructures. In the case of the Western
substructure this edge is the infalling front, while for the Eastern
substructure it is unclear whether the group is moving to the West,
as suggested by the appearance of the X-ray residuals, or
to the East, as suggested by the positions of NGC4911 and NGC4921
(Neumann et al. 2003).  

Overall, this strongly suggests that the k+a spectra, i.e. the
truncation of the star formation activity in these galaxies and
possibly the previous starburst, could be the result of an interaction
with the hot intracluster medium (ICM).  Hence, as far as Coma k+a galaxies
are concerned, there is suggestive evidence that the origin of the k+a
spectrum is a cluster-related and, in particular, an ICM-related
phenomenon that is closely connected with the dynamical state of the
cluster.  We note that the $\rm H\delta$ strength of the blue k+a
galaxies implies that star formation was truncated in these galaxies
on a short timescale, i.e. short compared to the k+a timescale of
1-1.5 Gyr. In fact, a slowly declining star formation activity such as
that envisaged if galaxies simply lost their gas halo reservoir when
becoming part of a group (``strangulation'', e.g. Bower \& Balogh
2003) is not able to produce such strong Balmer lines.

It is worth stressing the importance of isolating the {\it youngest}
k+a galaxies using their blue colors and strong equivalent widths. The
red k+a phase has a timescale that is comparable to the core crossing
time in a cluster like Coma, and any signature of the link between the
truncation of star formation and the location within a substructure is
thus erased in the older k+a's, while it is still detectable in the
youngest subsample of k+a's.

\begin{figure}[h]
\plotone{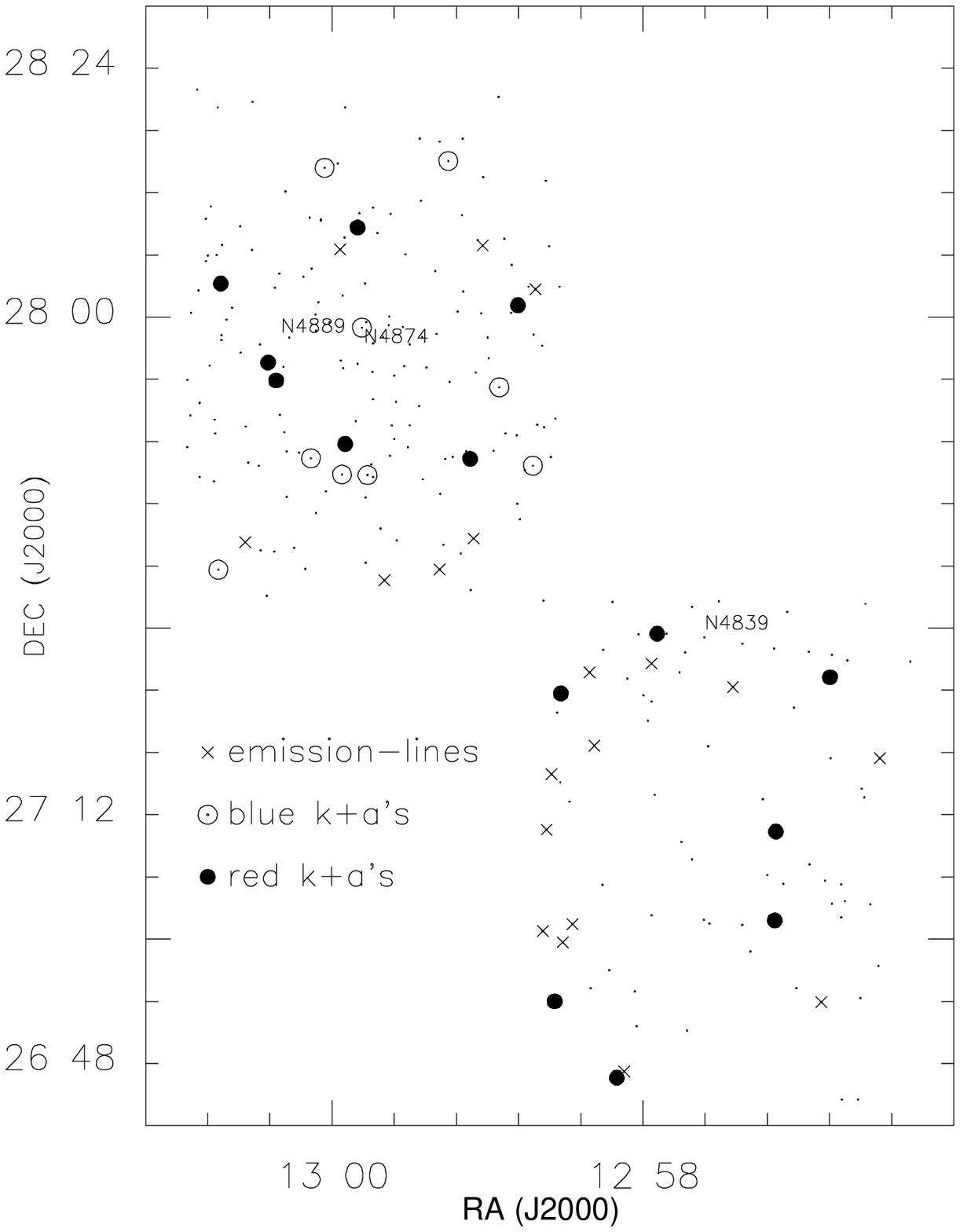}
\caption{Projected position on the sky of galaxies with passive spectra 
(small dots), red and blue k+a (large circles) and emission-line
(crosses) galaxies. The location of the three dominant galaxies
(N4874, N4889 and N4839) is labelled.}
\end{figure}

It is also instructive to note that looking for a spatial segregation
in the location of galaxies on the sky (Fig.~2) would not allow to
establish a correlation between
the star formation history of the k+a galaxies and the substructure.
The link with the dynamical history of Coma
appears evident only once a
detailed X-ray map reveals the complicated structure in the hot
intracluster gas.

\begin{figure}[h]
\plotone{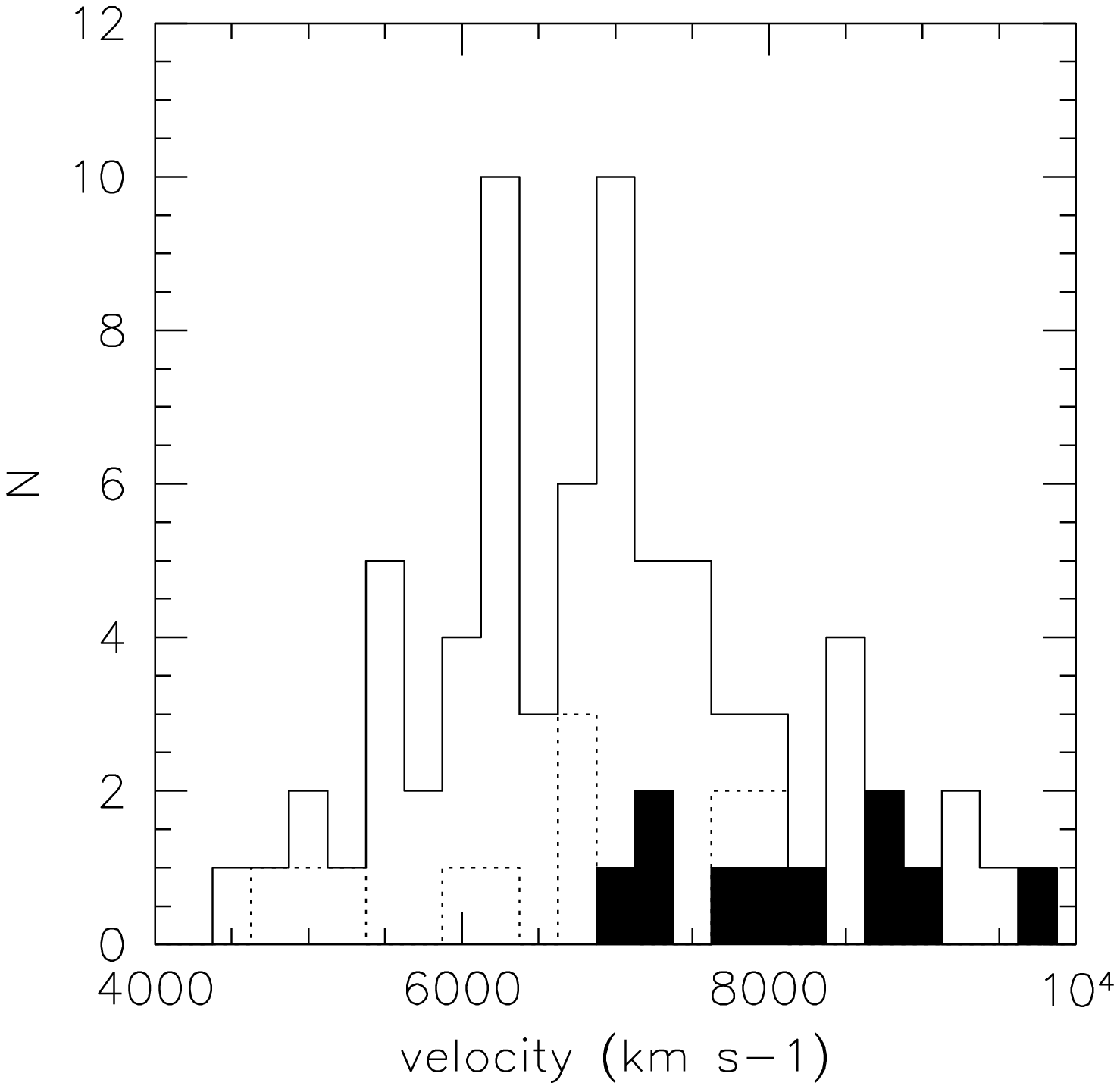}
\caption{Velocity histograms of all Coma dwarfs (solid line), strong blue k+a's
(filled histogram) and weak red k+a's (dashed line).}
\end{figure}

The blue k+a's do show, however, a radial velocity distribution that
is significantly different from that of the red k+a's and the global
Coma dwarf population, as shown in Fig.~3.  Their mean radial velocity
is 8120$\pm 709$ km $\rm s^{-1}$, with all but one at $v>7200$ km $\rm
s^{-1}$.  In contrast, both the red k+a's and all faint galaxies with
passive spectra have much lower mean velocities: 6992$\pm 761$ and
6854$\pm 244$ km $\rm s^{-1}$, respectively.  The blue k+a galaxies
have a relatively high velocity dispersion (1250 $\rm km \, s^{-1}$)
already indicating that they cannot be part of a single infalling
bound galaxy group of relatively small mass compared to the whole Coma
cluster.

Interestingly, also the (few) emission line galaxies in the central region 
of Coma seem to trace the
substructure: they are mostly found along
the edge of the Western structure (Fig.~4), with the exception of
\#15480 that is located just South of the Eastern structure.
Emission-line galaxies are known to be proportionally more numerous in
the South-West region than towards the central region of the cluster
(Fig.~4).

\begin{figure}[h]
\plotone{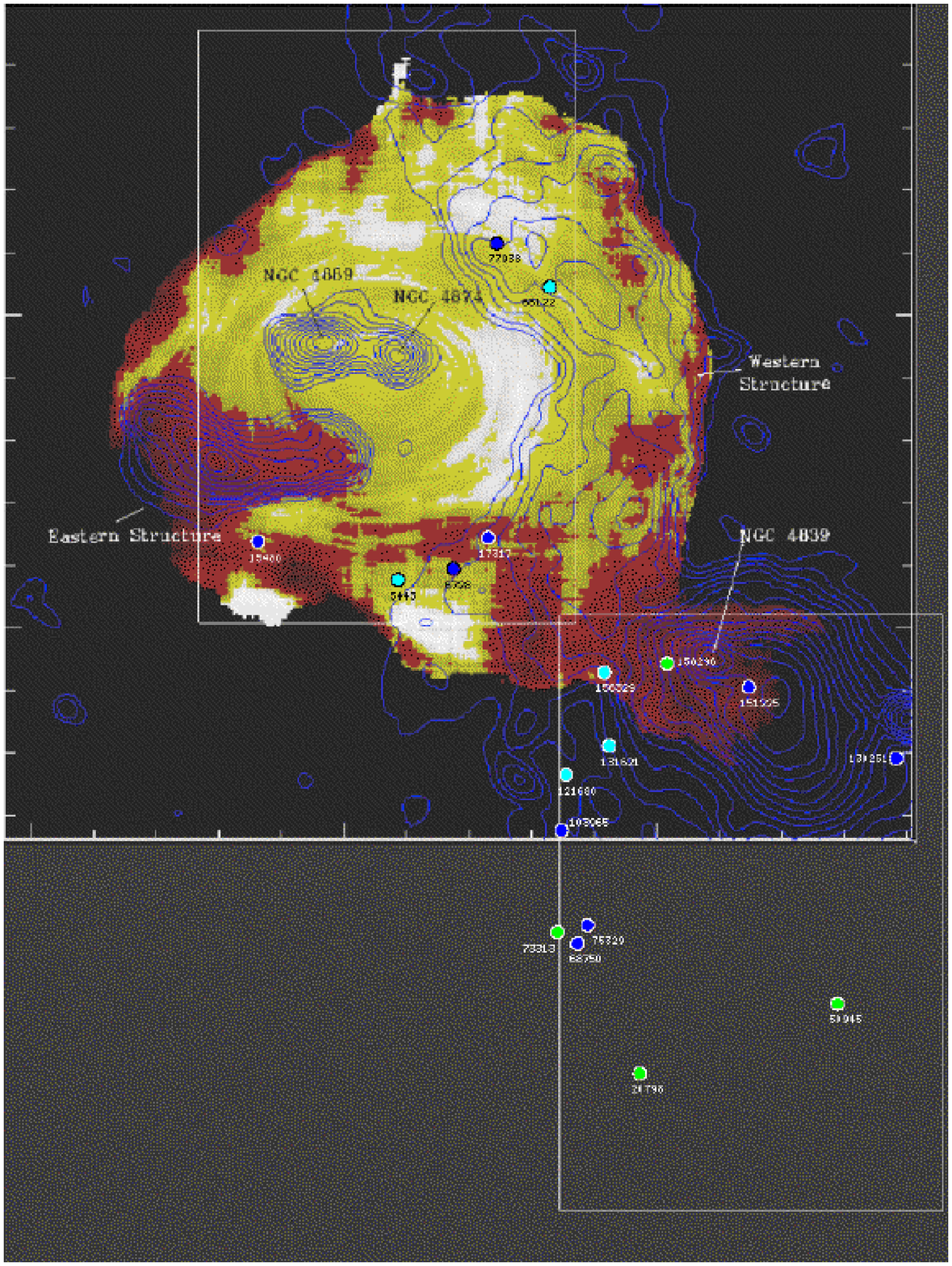}
\caption{Same as Fig.~1, but now showing the positions of emission-line 
galaxies instead of k+a's. 
Dots of different colors represent galaxies with different levels of
star formation activity, but all dots are associated with galaxies
with ongoing star formation.}
\end{figure}

\section{Conclusions}
The position of faint post-starburst galaxies in Coma relative
to X-ray substructure in the cluster strongly suggests that
the interruption of the star formation activity in these galaxies 
is a cluster-related phenomenon, most likely due to
the impact with the intracluster medium.
Post-starburst galaxies in Coma are much fainter than those
observed in distant clusters, a difference probably reflecting
a change in the galaxy populations infalling in clusters.

The dynamical state of the cluster and its recent history of accretion 
are clearly important factors in determining the galaxy properties observed.  
Identifying substructure, however, it is often hard, and no easily
measurable ``global'' cluster property (such as for example X-ray
luminosity or richness) is sufficient to characterize the cluster
dynamical status.  We have shown here an example in which only a 
detailed analysis of the X-ray residuals has revealed the link between
substructure and rapid changes in the star formation histories of
some of the cluster galaxies.

\acknowledgments{BMP thanks the IAU and the organizers of this Symposium 
for their kind invitation and for generously
supporting her participation with a IAU travel grant.}


\begin{references}
\reference Abraham, R.G., Smecker-Hane, T.A., Hutchings, J.B., Carlberg, R.G., Yee, H.K.C., Ellingson, E., Morris, S., Oke, J.B., Rigler, M., 1996, ApJ, 471, 694
\reference Bahcall, N.,A., 1977, ApJ, 218, L93
\reference Balogh, M.~L., Morris, S.~L., Yee, H. K. C., Carlberg, R. G., \& Ellingson, E., 1999, \apj, 527, 54
\reference Balogh, M.L., Bower, R.G., Smail, I., Ziegler, B.L., Davies, R.L., Gaztelu, A., Fritz, A., 2002a, MNRAS, 337, 256
\reference Balogh, M.L., Smail, I., Bower, R.G., Ziegler, B.L., Smith, G.P., Davies, R.L., Gaztelu, A., Kneib, J.-P., Ebeling, H., 2002b, ApJ, 566, 123
\reference Bower, R.G. \& Balogh, M.L., 2003
in Carnegie Observatories Astrophysics Series, Vol.3:
Clusters of Galaxies: Probes of Cosmological Structure and Galaxy
Evolution, eds. J.S. Mulchaey, A. Dressler and A. Oemler (Cambridge: 
Cambridge University Press), http://www.ociw.edu/ociw/symposia/series/symposium3/proceedings.html 
\reference Caldwell, N., Rose, J.A., Sharples, R.M., Ellis, R.S., Bower, R.G., 1993, AnJ, 106, 473
\reference Couch, W.J., Sharples, R.M., 1987,
MNRAS, 229, 423
\reference Dressler, A., 1980, ApJ, 236, 351
\reference Dressler, A., Gunn, J.E., 1992, ApJS, 78, 1
\reference Dressler, A., Oemler, A., Jr., Couch, W. J., Smail, I., Ellis, R. S., Barger, A., Butcher, H., Poggianti, B. M., Sharples, Ray M., 1997, ApJ, 490, 577
\reference Dressler, A., Smail, I., Poggianti, B. M., Butcher, H., Couch, W. J., Ellis, R. S., \& Oemler, A. 1999, \apjs, 122, 51
\reference Dwarakanath, K.S., Owen, F.N., 1999, AJ, 118, 625
\reference Edge, A.C., Stewart, G.C., 1991, MNRAS, 252, 428
\reference Ellingson, E., Lin, H., Yee, H. K. C., \& Carlberg, R. G. 2001, \apj, 547, 609
\reference Fisher, D., Fabricant, D., Franx, M., van Dokkum, P., 1998, ApJ, 498, 195
\reference Gomez, P.L., Nichol, R.C., Miller, C.J., et al., 2003, ApJ, 584, 210
\reference Lewis, I., Balogh, M., De Propris, R. et al., 2002, MNRAS, 334, 673
\reference Metevier, A.J., Romer, A.K., Ulmer, M. P. 2000, AJ, 119, 1090
\reference Miller, N.A., 2003, in Carnegie Observatories Astrophysics Series, vol.3: Clusters of Galaxies: Probes of Cosmological Structure and Galaxy Evolution (eds. J.S. Mulchaey, A. Dressler, and A. Oemler (Pasadena: Carnegie Observatories, http://www.ociw.edu/ociw/symposia/series/symposium3/proceedings.html)
\reference Miller, N.A, Owen, F.N., 2003, AJ, 125, 2427
\reference Mobasher, B., Bridges, T.\,J., Carter, D., Poggianti, B.\,M., et al., 2001. ApJS, 137, 279  
\reference Neumann, D.M., Lumb, D.H., Pratt, G.W., Briel, U.G., 2003, A\&A, 400, 811
\reference Oemler, A., 1974, ApJ, 194, 1
\reference Pimbblet, K.A., Smail, I., Kodama, T., Couch, W., Edge, A.C., Zabludoff, A.I., O'Hely, E., 2002, MNRAS, 331, 333
\reference Poggianti, B.M., Bridges, T.J., Komiyama, Y., Yagi, M., Carter, D., Mobasher, B., Okamura, S., Kashikawa, N., 2003, ApJ in press (P03, astro-ph 0309449)
\reference Postman, M., Geller, M.J., 1984, ApJ, 281, 95
\reference Smail, I., Edge, A.C., Ellis, R. S., Blandford, R.D., 1998, MNRAS, 293, 124
\reference Tran, K.-V.H., Franx, M., Illingworth, G., Kelson, D.D., van Dokkum, P., 2003, ApJ in press (astro-ph 0309460)
\reference Treu, T., Ellis, R.S., Kneib, J.-P., Dressler, A., Smail, I., Czoske, O., Oemler, A., Natarajan, P., 2003, ApJ, 591, 53
\reference van Gorkom, J., 2003, in Carnegie Observatories Astrophysics Series, vol.3: Clusters of Galaxies: Probes of Cosmological Structure and Galaxy Evolution (eds. J.S. Mulchaey, A. Dressler, and A. Oemler (Pasadena: Carnegie Observatories, http://www.ociw.edu/ociw/symposia/series/symposium3/proceedings.html)
\reference Wang, Q.D., Ulmer, M.P., 1997, MNRAS, 292, 920
\end{references}
\end{document}